# Tool-mediated HCI Modeling Instruction in a Campus-based Software Quality Course

Christos Katsanos[1,2], Michalis Xenos[3] and Nikolaos Tselios[4]

[1] Department of Informatics, Aristotle University of Thessaloniki, Thessaloniki, Greece
[2] HCI Group, Electrical and Computer Engineering Department, University of Patras, Patras, Greece
[3] Computer Engineering and Informatics Department, University of Patras, Patras, Greece
[4] Educational Sciences and Early Childhood Education Dept., University of Patras, Patras, Greece
ckatsanos@ece.upatras.gr, xenos@ceid.upatras.gr, nitse@ece.upatras.gr

**Abstract.** The Keystroke Level Model (KLM) and Fitts' Law constitute core teaching subjects in most HCI courses, as well as many courses on software design and evaluation. The KLM Form Analyzer (KLM-FA) has been introduced as a practitioner's tool to facilitate web form design and evaluation, based on these established HCI predictive models. It was also hypothesized that KLM-FA can also be used for educational purposes, since it provides step-by-step tracing of the KLM modeling for any web form filling task, according to various interaction strategies or users' characteristics. In our previous work, we found that KLM-FA supports teaching and learning of HCI modeling in the context of distance education. This paper reports a study investigating the learning effectiveness of KLM-FA in the context of campus-based higher education. Students of a software quality course completed a knowledge test after the lecture-based instruction (pre-test condition) and after being involved in a KLM-FA mediated learning activity (post-test condition). They also provided post-test ratings for their educational experience and the tool's usability. Results showed that KLM-FA can significantly improve learning of the HCI modeling. In addition, participating students rated their perceived educational experience as very satisfactory and the perceived usability of KLM-FA as good to excellent.

## 1 Introduction

The integration of core human computer interaction (HCI) concepts such as interface design and evaluation into the computer science/engineering curriculum is not well anticipated. Such an integration should balance effectively HCI theory instruction as well as hands-on experience. Nevertheless, at the end of the learning process the students should be able to effectively use HCI knowledge to design and evaluate soft-ware. As a result, teaching of independent, isolated conceptual entities without offering a coherent conceptual context to provide the student the ability to create meaningful associations and abstractions and subsequently apply the obtained knowledge to increase quality of interaction should not be considered as effective [1, 2].

The keystroke level model (KLM) [3] provides an accurate estimation of the time required to perform an interaction task for an expert and error-free modeled user. The designer can employ this model to choose among design alternatives and optimize a user interface in terms of task completion time. As a result, an efficient KLM learning activity should provide the opportunity to the student to: a) learn the basic aspects of the KLM model, b) to apply efficiently the KLM to calculate the time to complete a task with a given user interface, c) to understand the impact of specific design decisions on usability, and d) to understand the differences of alternative interaction strategies. Web form filling tasks provide a rather good learning context to introduce students to KLM modeling because people are typically experienced in performing themselves such tasks and form filling tasks are well-organized and executed in a serial manner.

A tool which automates the process of calculating the time required to complete a form may greatly assist the educational process. It can provide the ability to test alternative scenarios and to reflect upon specific design approaches without the need to carry out tedious and repetitive calculations. The Keystroke Level Model-Form Analyzer (KLM-FA)[1] has been introduced [4, 5], as a practitioner's tool to facilitate web form design and evaluation based on established HCI predictive models, namely the Keystroke Level Model (KLM) [3] and the Fitts' Law [6]. Such models constitute a core teaching subject of HCI courses, as well as of courses like software design, interaction design and software quality. In a previous work [7], it was found that KLM-FA can support teaching and learning of HCI modeling in the context of distance education. In this paper, we employ a pre-post research design in the context of traditional campus-based education. The tool is briefly presented hereinafter.

### 1.1 The KLM Form Analyzer Tool (KLM-FA)

KLM-FA is a Windows desktop application for web form design and evaluation. The tool automatically detects the form elements and presents them in an embedded web form preview. The user of the tool can select the form elements involved in the modeled task –they are auto-highlighted in the web form preview– and KLM-FA calculates the predicted time for task completion (see Fig. 1). The tool provides various profiles for modeled users depending on their age and typing skills following established KLM modeling conventions. In addition, KLM-FA can easily model various form filling interaction scenarios depending on input devices usage (keyboard, mouse) for filling form elements.

KLM-FA is based on a two-phase modeling of every user action on a web form element: one has to first *reach* the form element and next to *manipulate* it. A typical usage scenario of the tool is the following. First, the designer inputs a URL of a web form either stored online in a web server or locally in the designer's filesystem. Then, KLM-FA parses the webpage and identifies all forms and form elements. Subsequently, the designer selects all or some of the elements which are involved in the modeled user interaction task, specifies related parameters (e.g. whether the user is using the mouse to reach form elements) and may optionally change KLM modeling defaults (e.g. rules for placing mental operators).

---
[1] Available at http://klmformanalyzer.weebly.com

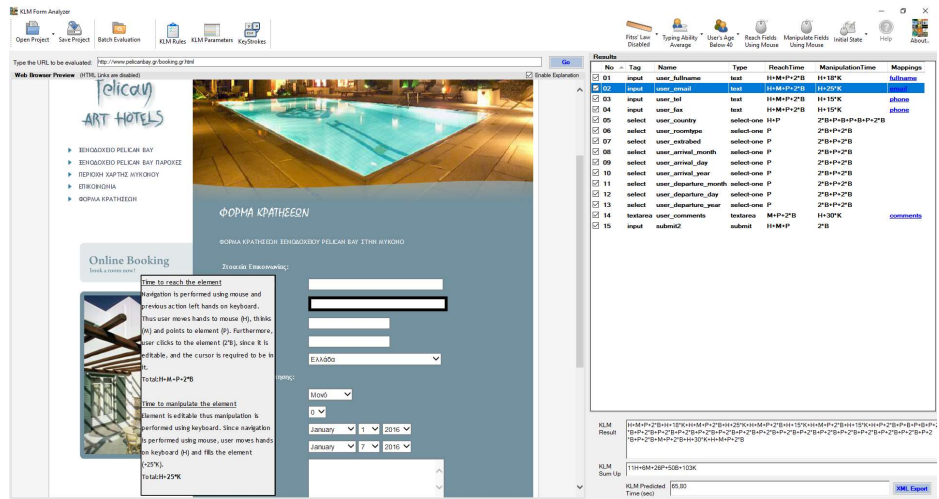

**Fig. 1.** The KLM-FA main user interface: an interactive web form preview shows the web form (left part) and an interactive table of modeling results shows KLM operators and time estimation per element and in total (right part). The "Enable Explanation" functionality displays a tooltip explaining the KLM modeling sequence next to the currently focused element.

Next, KLM-FA simulates the series of modeled user's actions to first reach each element and then manipulate it. In this context, KLM operators are being detected and furthermore Fitts' Law may be automatically applied for each of the simulated mouse movements to reach or manipulate an element. Then, a summarized result presents the series of simulated actions to complete the form filling task and estimates the required time to do so.

KLM-FA is useful for educational purposes, apart for professional design or evaluation practices, since it supports the following educational scenarios [7]:

- It provides *step-by-step KLM application*. The student can select a form element either from the web browser or from the sequential list and observe the specific actions to reach and manipulate the element. As a result, the student can distinguish what element of the form contributes to what extent in the total KLM calculation.

- It provides *concrete feedback* to the user. The KLM-FA user can alter some of the assumptions describing the modeled user (i.e., age, typing ability) and observe the difference.

- It provides the ability to *test alternative designs*. Thus, the user can investigate the design approaches which lead to an optimum result, in terms of time to complete the form-filling task.
- It provides the ability to *explore the impact of form elements' positioning and size*. By using Fitts' law to compute the time to point an element, the user may observe the impact of element's size and relative positioning into the form.
- It provides the ability to *test alternative form manipulation strategies* such as keyboard usage, mouse usage or a combination of them.

This paper reports a pre-post study that investigates the learning effectiveness of KLM-FA (post-test) compared to lecture-based instruction (pre-test) in the context of campus-based higher education. In specific, the research questions investigated by this study are the following:

- RQ1: Is there any effect of the KLM-FA mediated learning activity on students' learning performance in the context of campus-based higher education?
- RQ2: Were students with lower pretest score benefited from the KLM-FA mediated learning activity at least to the same extent as students with higher pretest score?
- RQ3: Did students find the KLM-FA a useful educational tool?
- RQ4: Did students find the KLM-FA a usable tool?

The rest of the paper is organized as follows. The method of the study is presented in the next section, followed by the findings which are analyzed according to the aforementioned research questions presented. Finally, the paper concludes with a discussion of the obtained findings as well as directions for future work.

## 2 Methodology

The study presented in this paper provides insights on the students learning gain (academic performance), as well as on the students' perceived educational experience. In addition, it provides the results of the students' assessment of the KLM-FA, the tool used to mediate the learning activity. A one group pre-post design was adopted.

The study took place in the context of campus-based classroom education and in specific in the course named "*CEID_NE5577: Software Quality Assurance and Standards*". This is an elective course offered to the students of the Computer Engineering and Informatics Department (CEID) at the University of Patras, during the first semester of their final (5th) year of studies. CEID is a 5-year B.Sc. degree with an Integrated M.Sc., corresponding to 300 European Credit Transfer System (ECTS) units.

The CEID_NE5577 course includes 14 lectures and 5 compulsory essays and offers 5 ECTS units to the students. The essays are graded and contribute to 50% of the final course grade, while a passing grade in all essays is a prerequisite for participating in the final exams for the other 50% of the course grade. Students are introduced to basic HCI concepts during lectures 2 to 5. These lectures include, inter alia, the presentation of the topics related to this study: The Human Information Processor model, the KLM and the Fitts' Law.

### 2.1 Participants

There were 108 students enrolled in the CEID_NE5577 course. However, only 22 of them attended the lectures, participated in all essays and successfully completed the course. These 22 students were aged from 21 to 26 (mean age = 22.1, SD = 0.97) and 7 were female. Although the lecture attendance in CEID is not obligatory, this course had a high attendance rate with approximately 17 out of the 22 students attending each lecture (mean number of students per lecture = 17.2, SD = 3.8) and an equally high participation on the course section of the eClass, which is the University of Patras Learning Management System (LMS).

The eClass LMS offers means of asynchronous communication (e-messages and e-fora) and the appropriate infrastructure for receiving, submitting and grading the essays. A total of 287 students' messages were recorded in the eClass (mean number of messages per student = 13.0, SD = 5.2) which is interpreted as a very active online presence, especially since CEID_NE5577 is a campus-based and not a distance course. All communication between students and the professor was through eClass and all essays were submitted and graded on the eClass.

### 2.2 Materials

The KLM and Fitts' Law are part of the second essay, in which the students were required to demonstrate their knowledge, by solving some modeling problems in interaction design using KLM-FA. The students had 15 days to complete the essay.

Regarding the KLM part of the essay, students were provided with the KLM-FA tool and a short (6m23s) video available on YouTube further explaining the use of the tool. They were asked to use KLM-FA to evaluate signup forms of diverse complexity –from easy signup forms of hotels as shown in Fig.1, to more complex forms of online services– using various parameters of the tool (i.e. user ages, typing skill), as well as using Fitts' Law to model the pointing device movement time. Students were also asked to conduct a "pen-and-paper" KLM modeling for a non-working prototype of a, rather complex, form.

All students were asked to complete a knowledge test (pre-test) after the end of the lectures and before downloading their essay description from the course LMS. They had access to their essay description only after completing the knowledge test. The knowledge test included all issues related to KLM and comprised of 14 multiple-choice questions with four answer options each. The same test (post-test) was offered to them after they had submitted their essays and completing it was required to formally finalize their essays into the course LMS. They were also asked to complete three additional scales as part of the post-test: a) a 5-items scale rating their educational experience with KLM-FA from 1 to 5, b) the standardized System Usability

Scale [8] provided in participants' native language [9, 10], and c) the 7-point adjective rating question [11] with wordings from "worst-imaginable" to "best-imaginable".

### 2.3 Procedures

The learning material, including the lectures slides, KLM-FA demo video and the essay description were announced on the eClass. All the questionnaires were also incorporated in the same LMS. After the deadline, the 22 submitted essays were graded by the professor and indicative solutions and results were uploaded in eClass. Finally, each student received personal remarks on their submitted essay.

The collected data were organized and preprocessed using Microsoft Excel 365 ProPlus and were analyzed using IBM SPSS Statistics v20.0. The materials offered to students using the eClass LMS were available to all enrolled students until the end of the semester.

## 3   Results

First, reliability analysis was conducted for the questionnaires used in the study. To this end, the Cronbach's alpha measure of internal consistency was used [12]. The knowledge test had low reliability with all items included; Cronbach's α = 0.402, N = 14 items. This was attributed to an ambiguity in the way the content related to two questions was presented in the lecture slides and in the references at the course book. Removing these two questions resulted in a scale with adequate reliability; Cronbach's α = 0.782, N = 12 items. The educational experience scale had very good internal consistency; Cronbach's α = 0.873, N = 5 items. SUS is a standardized scale [9, 10, 13–15] and had also adequate reliability for our dataset; Cronbach's α = 0.760, N = 10 items.

Following the rationale reported in [16], we produced a composite variable for the normalized learning gain defined as the difference between posttest score and pretest score ("observed gain" [16]) divided by the difference between the max possible score and the pretest score ("amount of possible learning that could be achieved" [16]).

**Table 1.** Descriptive statistics of the dependent variables in this study. Sample size N = 22 university students in campus-based education.

| Variable | M | Mdn | SD | 95% CI |
|---|---|---|---|---|
| Pretest score (0-100) | 62.9 | 66.7 | 15.4 | [56.0, 69.7] |
| Posttest score (0-100) | 72.0 | 75.0 | 13.0 | [66.2, 77.7] |
| Normalized learning gain (%) | 19.1 | 18.3 | 37.0 | [2.7, 35.5] |
| KLM-FA educational experience rating (1-5) | 4.0 | 4.0 | 0.8 | [3.7, 4.3] |
| SUS score for KLM-FA (0-100) | 82.0 | 85.0 | 10.2 | [77.6, 86.6] |
| Usability adjective rating for KLM-FA (1-7) | 5.3 | 5.0 | 0.8 | [5.0, 5.7] |

Table 1 presents descriptive statistics of the dependent variables measured in this study. In all subsequent statistical analyses, the effect size r was calculated according to the formulas reported in [17].

### 3.1 RQ1: Learning Performance and KLM-FA

A two-tailed dependent samples t-test showed that the difference between students' pretest and posttest scores in the knowledge test was statistically significant; $t(21) = 2.890$, $p = 0.009$, $r = 0.533$. This large observed effect size [18] demonstrates the learning effectiveness of KLM-FA in the context of campus-based higher education. A parametric test was used, because Shapiro-Wilk analysis found that the distribution of the differences in the posttest and pretest scores did not deviate significantly from a normal distribution; $W(22) = 0.939$, $p = 0.187$.

However, a Kolmogorov-Smirnov test of normality showed that this distribution of differences in scores was significantly non-normal; $D(22) = 0.202$, $p = 0.02$. Although the Shapiro-Wilk test is more reliable for small sample sizes [17], we also conducted a non-parametric analysis of the same data due to the contradicting results in the normality tests. In agreement to the parametric test, a two-tailed Wilcoxon signed rank test also found that students improved significantly their scores in the knowledge test after using KLM-FA; $z = 2.601$, $p = 0.009$, $r = 0.390$.

### 3.2 RQ2: Learning Gain for Students with Low and High Pretest Score

A median split analysis was conducted to investigate whether students with low initial performance achieved higher learning gain compared to students with high initial performance.

To this end, students with pretest score lower to the median (Mdn = 66.7) were assigned in the low pretest performance condition (N = 10), whereas the rest were assigned in the high pretest performance condition (N = 12). Table 2 presents pretest score, posttest score and normalized learning gain grouped by students' initial performance.

**Table 2.** Descriptive statistics of study variables grouped by students' initial performance.

| Group | Variable | M | Mdn | SD | 95% CI |
| --- | --- | --- | --- | --- | --- |
| Low initial score | Pretest score (0-100) | 48.1 | 50.0 | 9.1 | [41.1, 55.1] |
| High initial score | Pretest score (0-100) | 74.2 | 75.0 | 8.7 | [68.4, 80.1] |
| Low initial score | Posttest score (0-100) | 66.7 | 66.7 | 9.3 | [59.5, 73.8] |
| High initial score | Posttest score (0-100) | 80.3 | 83.3 | 9.3 | [74.0, 86.6] |
| Low initial score | Normalized learning gain (%) | 34.0 | 40.0 | 21.1 | [17.7, 50.2] |
| High initial score | Normalized learning gain (%) | 21.2 | 0.0 | 27.2 | [2.9, 39.5] |

A two-tailed independent samples t-test found an effect of student's initial performance on their post-test score; $t(20) = 2.589$, $p = 0.018$. A parametric test was used, because both the assumptions of normality and homogeneity of variance were not violated; Shapiro-Wilk tests, p > 0.05 and Levene's test, F(1,20) = 0.157, p = 0.696 respectively. In addition, a two-tailed Man-Whitney U test found no effect of students' initial performance on their normalized learning gain; z = 1.292, p = 0.197. A non-parametric test was selected, because the assumption of normality was violated for the high initial performance group; W(12) = 0.787, p = 0.006.

These results tend to support that students of lower initial performance improved significantly more their posttest score compared to students of higher initial performance. However, students of both low and high initial performance were equally benefited from the KLM-FA activity in terms of their normalized learning gain.

### 3.3 RQ3: Educational Experience with KLM-FA

Participating students rated their learning experience with KLM-FA in the post-test questionnaire. Table 3 presents descriptive statistics of these ratings per question and overall.

**Table 3.** Descriptive statistics of students' self-reported ratings of their educational experience with KLM-FA.

| Question (1: strongly disagree; 5: strongly agree) | M | Mdn | SD | 95% CI |
|---|---|---|---|---|
| Q1. The KLM-FA helped me to understand the KLM model and Fitts' Law. | 3.9 | 4.0 | 0.9 | [3.5, 4.3] |
| Q2. During the activity, I am satisfied with my learning progress and effectiveness. | 3.7 | 4.0 | 0.9 | [3.3, 4.1] |
| Q3. I think that KLM-FA is useful as an educational tool. | 4.1 | 4.0 | 0.9 | [3.7, 4.5] |
| Q4. I would recommend KLM-FA to a colleague or friend who wants to learn the KLM and Fitts' Law. | 4.0 | 4.0 | 1.0 | [3.5, 4.4] |
| Q5. I would recommend KLM-FA to a colleague or friend who wants to learn how to design web forms or evaluate their usability. | 4.4 | 5.0 | 0.9 | [4.0, 4.8] |
| Overall scale (Cronbach's α=0.873) | 4.0 | 4.0 | 0.8 | [3.7, 4.3] |

Participants self-reported ratings about their learning experience with KLM-FA were rather high (M = 4.0, SD = 0.8). In specific, students agreed that KLM-FA helped them to understand the KLM model and Fitts' Law (M = 3.9, SD = 0.9) and that their perceived learning progress and effectiveness was satisfactory (M = 3.7, SD = 0.9). Students also found KLM-FA to be a useful educational tool (M = 4.1, SD = 0.9). The collected data suggest that they would probably recommend KLM-FA to colleagues or friends who want to be educated on established HCI models (M = 4.0, SD = 1.0) and web form design (M = 4.4, SD = 0.9). These results are in agreement with perceived educational experience ratings provided for KLM-FA by distance education university students in our previous work [7].

Correlation analysis found that students' education experience ratings and their grade in the KLM essay were not significantly associated; $r_s$=-0.360, p=0.100. Spear-man's coefficient was used, because the assumption of normality was violated by both variables; W(22)=0.903, p=0.035 and W(22)=0.828, p=0.001 respectively. This finding provides tentative support that students agreed that their educational experience was good without considering their academic performance in the essay.

In addition, students provided rather positive comments for their educational experience with KLM-FA in a related open-ended question of the post-test questionnaire. For instance, one student mentioned that "*It [KLM-FA] helped me understand the complexity of web forms design and factors that affect their usability*" and another mentioned that "*The 'Enable Explanation' feature helped me understand the calculation of the times to reach a field and to manipulate a field*".

In sum, these findings demonstrate that KLM-FA was perceived by the students as a useful educational tool that fits well into the educational context of the campus-based software quality course.

### 3.4 RQ4: Perceived Usability of KLM-FA

After interacting with KLM-FA, the participating students completed the SUS questionnaire and the adjective rating scale, both measures of a system's perceived usability.

KLM-FA received a mean SUS score of 82.0 (SD = 10.2). According to a dataset of nearly 1000 SUS surveys [11], this means that students found KLM-FA as "Good to Excellent" (SUS score from 71.4 to 85.5) in terms of perceived usability. Students' usability adjective ratings were also rather high (M=5.3, SD=0.8), confirming that KLM-FA was perceived from "Good" (corresponds to 5) to "Excellent" (corresponds to 6). In a previous study with distance education students [7], KLM-FA was also perceived as "Good to Excellent" with a mean SUS score of 73.6 (SD = 13.2) and similar usability adjective ratings (M=5.4, SD=0.8).

In the post-test questionnaire, students were also offered the chance to write down three positive and three negative characteristics regarding the KLM-FA tool. A total of 41 positive and 16 negative characteristics were reported respectively. Thematic analysis of participants' answers resulted in groupings. KLM-FA was found to be a usable and easy to use tool (8 students), provide quick and accurate results (8 students), educationally valuable (8 students), flexible and parametrizable (7), simple and understandable (6) and useful for Fitts' Law calculations (3). By contrast, students reported that KLM-FA needs: a) improvements in user interaction (6 students), such as adding shortcuts for frequently used functionality, b) changes in KLM modeling (5 students), such as modeling user errors, c) enrichment in modeling explanation (3 students), and d) additional flexibility by not requiring an implemented DHTML form (2 students). Finally, the most frequently-mentioned students' general suggestion for further KLM-FA improvement, as mentioned in a separate question, was to make it available in different operating systems other than Windows (3 students).

All in all, these findings demonstrate that KLM-FA was perceived as a usable tool by the students of the software quality course.

## 4      Discussion and Conclusions

The goal of this paper was to examine the effectiveness of a tool mediated learning activity in the context of a campus-based HCI modeling instruction. A one group pre-posttest design was adopted. Students improved significantly their scores in the knowledge assessment test after using KLM-FA, the tool that mediated the learning activity. Their performance was significantly improved and jumped, on average from 62.9% to 72%. Also, there were no significant differences in the students' learning gain of different initial performance (low versus high initial performance). The tool which has been adopted (KLM-FA) was perceived as a useful and usable tool by the students regardless of their academic performance. These findings are in agreement with our previous work both on using KLM-FA in the context of distance education [7] and implementing other tool-mediated activities [19, 20] to support HCI instruction.

However, this study is not without limitations. First, the sample was rather small and therefore the confidence interval for the results is rather wide. Moreover, an experiment with a control group using paper and pencil and a treatment group using the KLM-FA tool will help us to examine in more detail the exact learning phenomena taking place during the activity. In addition, utilization of learning analytics methods [21] to examine the students' low level actions and their possible relation with the learning outcome will provide a deeper understanding of the way the students benefited from the tool. Furthermore, exploration of other instructional design approaches which adopt technologies such as wikis [22, 23], combined with the introduction of KLM-FA in the HCI modeling instruction constitute an additional future goal. Finally, we also plan to conduct studies that monitor students' fixations [24, 25] and other physiological signals, such as skin conductance which is a reliable indicator of stress [26–29], in an attempt to evaluate and further improve the KLM-FA user experience.